\documentclass[a4paper,fleqn,usenatbib]{mnras}

\usepackage{amsmath}
\usepackage{natbib,twoopt}
\usepackage{newtxtext,newtxmath}

\usepackage[T1]{fontenc}
\usepackage{ae,aecompl}

\usepackage{graphicx}	
\usepackage{amsmath}	

\title[XMM and NuSTAR spectroscopy of MCG-5-23-16]{A remarkably stable accretion disc in the Seyfert galaxy MCG-5-23-16}

\author[R. Serafinelli et al.]{Roberto Serafinelli,$^{1}$\thanks{E-mail: roberto.serafinelli@inaf.it}, Andrea Marinucci$^{2}$, Alessandra De Rosa$^{1}$, Stefano Bianchi$^{3}$, 
\newauthor Riccardo Middei$^{4,5}$, Giorgio Matt$^{3}$, James N. Reeves$^{6,7}$, Valentina Braito$^{7,6,8}$, 
\newauthor Francesco Tombesi$^{9,5,10,11,12}$, Vittoria E. Gianolli$^{3,13}$, Adam Ingram$^{14}$, Fr\'ed\'eric Marin$^{15}$, 
\newauthor Pierre-Olivier Petrucci$^{13}$, Daniele Tagliacozzo$^{3}$, Francesco Ursini$^{3}$\\
\\
$^{1}$INAF - Istituto di Astrofisica e Planetologia Spaziali, Via del Fosso del Cavaliere 100, 00133, Roma, Italy\\
$^{2}$ASI - Agenzia Spaziale Italiana, Via del Politecnico snc, 000133, Roma, Italy\\
$^{3}$Dipartimento di Matematica e Fisica, Università degli Studi Roma Tre, Via della Vasca Navale 84, 00146, Roma, Italy\\
$^{4}$Space Science Data Center, Agenzia Spaziale Italiana, Via del Politecnico snc, 00133, Roma, Italy\\
$^{5}$INAF - Osservatorio Astronomico di Roma, Via Frascati 33, 00078, Monte Porzio Catone (Roma), Italy\\
$^{6}$Department of Physics, Institute for Astrophysics and Computational Sciences, The Catholic University of America, Washington,
DC 20064, USA\\
$^{7}$INAF - Osservatorio Astronomico di Brera, Via Bianchi 46, 23807, Merate (LC), Italy\\
$^{8}$Dipartimento di Fisica, Università di Trento, Via Sommarive 14, Trento 38123, Italy\\
$^{9}$Dipartimento di Fisica, Università degli Studi di Roma Tor Vergata, Via della Ricerca Scientifica 1, 00133, Roma, Italy\\
$^{10}$INFN - Istituto Nazionale di Fisica Nucleare, Sezione di Roma "Tor Vergata", Via della Ricerca Scientifica 1, 00133, Roma, Italy\\
$^{11}$Department of Astronomy, University of Maryland, College Park, MD 20742, USA\\
$^{12}$NASA Goddard Space Flight Center, Code 662, Greenbelt, MD 20771, USA\\
$^{13}$Université Grenoble Alpes, CNRS, IPAG, 38000 Grenoble, France\\
$^{14}$School of Mathematics, Statistics, and Physics, Newcastle University, Newcastle Upon Tyne NE1 7RU, United Kingdom\\
$^{15}$Université de Strasbourg, CNRS, Observatoire Astrononomique de Strasbourg, UMR 7550, 67000 Strasbourg, France\\
}

\date{Accepted XXX. Received YYY; in original form ZZZ}

\pubyear{2023}

\begin{document}
\label{firstpage}
\pagerange{\pageref{firstpage}--\pageref{lastpage}}
\maketitle

\begin{abstract}
MCG-5-23-16 is a Seyfert 1.9 galaxy at redshift $z=0.00849$. We analyse here the X-ray spectra obtained with {\it XMM-Newton} and {\it NuSTAR} data, which are the first contemporaneous observations with these two X-ray telescopes. Two reflection features, producing a narrow core and a broad component of the Fe K$\alpha$, are clearly detected in the data. The analysis of the broad iron line shows evidence of a truncated disc with inner radius $R_{\rm in}=40^{+23}_{-16}$ $R_g$ and an inclination of $41^{+9}_{-10}$ $^\circ$. The high quality of the {\it NuSTAR} observations allows us to measure a high energy cut-off at $E_{\rm cut}=131^{+10}_{-9}$ keV. We also analyse the RGS spectrum, finding that the soft X-ray emission is produced by two photoionised plasma emission regions, with different ionisation parameters and similar column densities. Remarkably, the source only shows moderate continuum flux variability, keeping the spectral shape roughly constant in a time scale of $\sim20$ years.\\

\end{abstract}




\section{Introduction}

The X-ray emission of active galactic nuclei (AGN) is thought to be produced by Comptonisation of UV seed photons coming from an accretion disc around a supermassive black hole (SMBH), by a hot corona, located in the innermost region around the SMBH \citep[e.g.,][]{haardt91,haardt93}. The X-ray spectrum of an AGN typically assumes the shape of a power law, characterised by a photon index $\Gamma$, up to a characteristic cut-off energy $E_{\rm cut}$, where the power law breaks. Both $\Gamma$ and $E_{\rm cut}$ are related to the physical properties of the corona, such as the electron temperature $kT_e$ and the optical depth $\tau_e$ \citep[e.g., ][]{shapiro76,sunyaev80,lightman87,petrucci01,middei19}. However, these relations are strongly dependent on the geometry of the corona, which is still largely unknown.\\
\indent Several measurements of the cut-off energy and/or the coronal temperature have been obtained with high-energy instruments such as {\it Beppo-SAX} \citep[e.g.,][]{dadina07}, {\it INTEGRAL} \citep{derosa12} and {\it Swift}-BAT \citep[e.g.,][]{ricci17}, but the launch of {\it NuSTAR} in 2013, with its unprecedented sensitivity in the $E>10$ keV energy band has revolutionised the field, providing high-precision cut-off energy measurements for dozens of AGN \citep[e.g.,][]{fabian15,fabian17}. The availability of high-quality measurements of the cut-off energy or coronal temperature, even for absorbed sources \citep[e.g.,][]{balokovic20,middei21,serafinelli23}, allowed several authors to study the relation between the cut-off energy and physical parameters of the AGN, finding that it is not dependent on the black hole mass $M_{\rm BH}$ or the accretion rate $L/L_{\rm Edd}$ \citep[e.g.,][Serafinelli et al., in prep.]{tortosa18,kamraj22}.\\
\indent However, the values of $kT_e$ and $\tau_e$ are degenerate with the cut-off energy and photon index, as spectroscopy alone is not able to distinguish between different geometries \citep[e.g., ][]{middei19,ursini22}, such as slab-like \citep[e.g.,][]{haardt91}, spherical \citep[e.g.,][]{frontera03} and lamp-post \citep[e.g.,][]{miniutti04}. Recently, the first coronal X-ray polarisation measurement has been obtained with the {\it Imaging X-ray Polarization Explorer} ({\it IXPE}) in the Seyfert galaxy NGC 4151 \citep{gianolli23}. The favoured geometry for the hot corona is a slab or a wedge distributed over the accretion disc \citep[see][for a description of the latter]{poutanen18}.\\
\indent MCG-5-23-16 is a Seyfert 1.9 galaxy \citep{veron80} at redshift $z=0.00849$. Due to its high X-ray flux ($F_{2-10\;{\rm keV}}\simeq8\times10^{-11}$ erg cm$^{-2}$ s$^{-1}$) the object was observed several times by many X-ray facilities in the last few decades.\\ 
\indent The X-ray spectrum of this AGN is characterised by moderate neutral absorption ($N_{\rm H}\simeq1.3\times10^{22}$ cm$^{-2}$) and by the presence of a soft excess and complex Fe K$\alpha$ emission. Evidence of a broad Fe K$\alpha$ line on this source was found with {\it ASCA} \citep{weaver97} and later confirmed by {\it XMM-Newton} observations \citep{dewangan03,balestra04}, pointing to the presence of two reflectors, one responsible for a narrow core at $6.4$ keV and one responsible for the broad wing, likely originated very close to the black hole.\\ 
\indent The broad line was explored with {\it Suzaku} \citep{reeves07}, and it was found that it is originated from a disc with inner radius of $\sim40R_g$, where $R_g=GM/c^2$ is the gravitational radius, from the central black hole, and inclination of $\sim50^\circ$. \cite{braito07} analysed a long simultaneous {\it XMM-Newton} ($131$ ks) and {\it Chandra} ($50$ ks) observation, which was able to confirm these results on the broad Fe K$\alpha$ line. Additionally, these data unveiled the presence of an absorption feature at $7.7$ keV, hinting at the possible presence of ionised iron outflowing at $\sim0.1c$, i.e. an ultra-fast outflow \citep[UFO, e.g.,][]{pounds03,tombesi10,serafinelli19,matzeu23}. Moreover, the RGS data allowed the detection of emission lines from ionised gas, pointing to a narrow-line region origin of the soft excess.\\
\indent The source was also observed by {\it NuSTAR}, which allowed the first accurate measurement of the cut-off energy \citep[$E\simeq120$ keV,][]{balokovic15}, and also hinting at its possible variability \citep{zoghbi17}. At the same time, {\it NuSTAR} confirmed the line parameters, such as the disc inner radius and inclination.\\
\indent Finally, MCG-5-23-16 was targeted by two pointings in 2022 with {\it IXPE} of $486$ ks \citep{marinucci22} and $642$ ks \citep{tagliacozzo23}, respectively. The combined observations yielded an upper limit $\Pi_{\rm max}=3.2$ on the polarization degree, in the $2-8$ keV energy band.\\
\indent Here, we present the analysis of the {\it XMM-Newton} and {\it NuSTAR} observations, the latter being taken together with the two {\it IXPE} pointings. In Sect.\ref{sec:data} we report on the data and the data reduction techniques that we adopted. We describe the spectral analysis, first performed in the $3-79$ keV band and then expanded to the broad band ($E=0.3-79$ keV) in Sect.~\ref{sec:spectral}. We summarise the results in Sect.~\ref{sec:theend}. Throughout the paper, we adopt a standard flat $\Lambda$CDM cosmology with $H_0=70$ km s$^{-1}$ Mpc$^{-1}$, $\Omega_m=0.3$ and $\Omega_\Lambda=0.7$.

\section{Data reduction}
\label{sec:data}

\begin{table}
    \centering
    \begin{tabular}{clccc}
    \hline
     Epoch & Instrument    & OBSID & Date & Exposure (s)\\
     \hline
     1 & {\it XMM-Newton}    & 0890670101 & 2022-05-21 & 92700\\
     1 & {\it NuSTAR} & 	60701014002 & 2022-05-21 & 83676\\
     2 & {\it NuSTAR} & 90801630002 & 2022-11-11 & 85743\\
     \hline
    \end{tabular}
    \caption{The data used in this paper. One of the {\it NuSTAR} observations is contemporaneous with {\it XMM-Newton} at Epoch 1. The {\it XMM-Newton} exposure is only the one from EPIC-pn, while the {\it NuSTAR} exposure must be read as per FPM detector.}
    \label{tab:data}
\end{table}

The data analysed here consists of contemporaneous {\it XMM-Newton} (OBSID 0890670101, 92 ks elapsed exposure time, 56 ks net) and {\it NuSTAR} observation (OBSID 60701014002, 83 ks), complementary to the {\it IXPE} observations analysed in \cite{marinucci22}, taken in May 2022 (Epoch 1). Additionally, the data set includes a further {\it NuSTAR} pointing (OBSID 90801630002, 85 ks), contemporaneous with the {\it IXPE} observation reported in \cite{tagliacozzo23}, taken in November 2022 (Epoch 2). The data are summarised in Table~\ref{tab:data}.\\
\indent We extract the event list from the EPIC-pn camera using the standard System Analysis Software ({\scriptsize SAS}) version 20.0.0 tool {\scriptsize EPPROC}, and we remove the flaring events \cite[e.g.,][]{deluca04} adopting an appropriate filtering to veto all those times where the observation is affected by flaring. We also extract the MOS spectra using the tool {\scriptsize EMPROC}, but they are severely affected by pile-up \cite[e.g.,][]{ballet99} and therefore they are not analysed here. The EPIC-pn spectrum is extracted by selecting a $40''$ radius region on the source, while the background is extracted on a source-free region of the same size. The response matrix and ancillary files are extracted with the standard tools {\scriptsize RMFGEN} and {\scriptsize ARFGEN}, respectively. We corrected the effective area with {\scriptsize APPLYABSFLUXCORR}, for a better agreement with {\it NuSTAR} data. The spectrum is grouped at a minimum of $100$ counts per energy bin. We consider the energy range $E=0.5-10$ keV for the EPIC-pn spectrum. We also reduce the RGS1 and RGS2 spectra using the {\scriptsize RGSPROC} task in {\scriptsize SAS}, with response matrices produced with {\scriptsize RGSRMFGEN}. We combine the two RGS spectra using {\scriptsize RGSCOMBINE} and we consider the energy band $E=0.5-2$ keV, i.e. $\lambda=6-25$ {\AA}.\\
\indent The {\it NuSTAR} data are reduced by using the {\scriptsize HEASOFT} v6.30 task {\scriptsize NUPIPELINE}, from the {\scriptsize NUSTARDAS} software package. We used {\scriptsize CALDB} calibration files updated as of August 29th, 2022. For each observation, we extract spectra from the two Focal Plane Modules A and B (FPMA and FPMB) by selecting a region with $60$'' radius around the source, and two source-free regions with $40$'' radius for the background. Both the detectors FPMA and FPMB are grouped at a minimum of $100$ counts per energy bin, for each observation. We consider the energy band $E=3-79$ keV for these data sets.

\section{Spectral analysis}
\label{sec:spectral}

We use the software {\scriptsize XSPEC} v12.12.0 \citep{arnaud96} to perform all spectral fits in this paper. The errors are reported at $90\%$ confidence level, corresponding to $\Delta\chi^2=2.71$. All models include a cross-calibration constant between the FPM modules and EPIC-pn, and in all our models is well fitted by the value $C_{\rm FPMA/pn}=1.39\pm0.01$ and $C_{\rm FPMB/pn}=1.42\pm0.01$. We assumed that these constants do not change between the two {\it NuSTAR} observations, while we let the normalisations free to take into account flux differences. For each epoch, the two FPM modules are fitted separately, but keeping each parameter of the fit tied together with the exception of the calibration constant. Following recent works \citep[e.g.,][]{gianolli23,ingram23}, we shift models that include lines with fixed centroid energies using the {\scriptsize XSPEC} model {\scriptsize VASHIFT}, to counter the possible presence of calibration issues. We leave the velocity free in the EPIC-pn data, and in the two FPMA spectra, with the shift velocities in the two FPMB observations tied to the simultaneous FPMA. In all models, we find a shift of $v_{\rm shift,XMM}=2200\pm300$ km s$^{-1}$ and $v_{\rm shift,NuSTAR,1}=4300\pm900$ km s$^{-1}$ and $v_{\rm shift,NuSTAR,2}=5500\pm1000$ km s$^{-1}$ for EPIC-pn and the FPM modules in Epochs 1 and 2, respectively. Initially, we assume that the spectral shape does not change between Epoch 1 and 2, therefore we keep $\Gamma$ and $E_{\rm cut}$ tied. The possible variability of these two parameters are discussed in Sect.~\ref{sec:discocont}. \\

\subsection{Hard band spectral analysis}
\label{sec:hardband}

\begin{figure}
\includegraphics[scale=0.35]{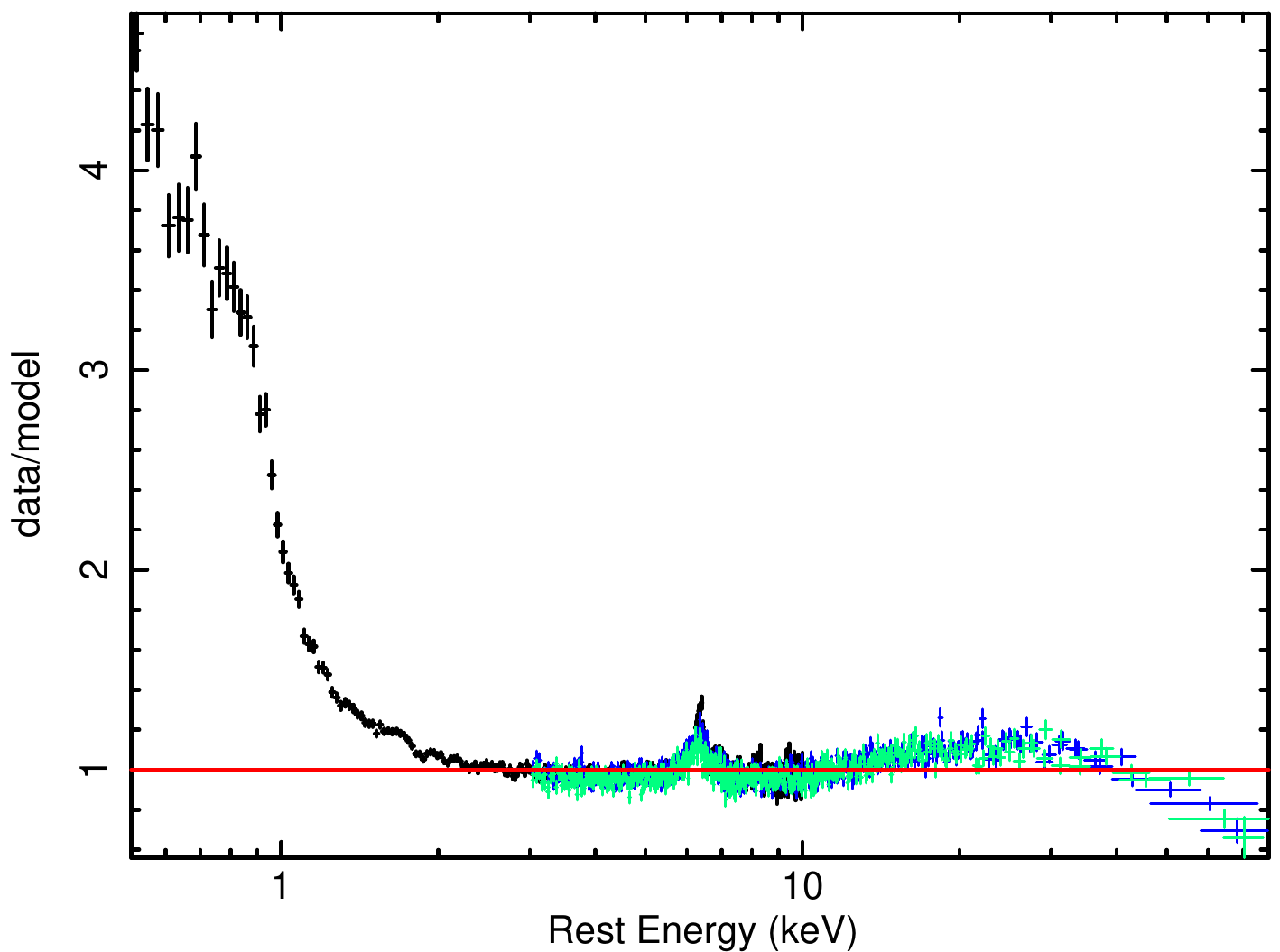}\\
\includegraphics[scale=0.35]{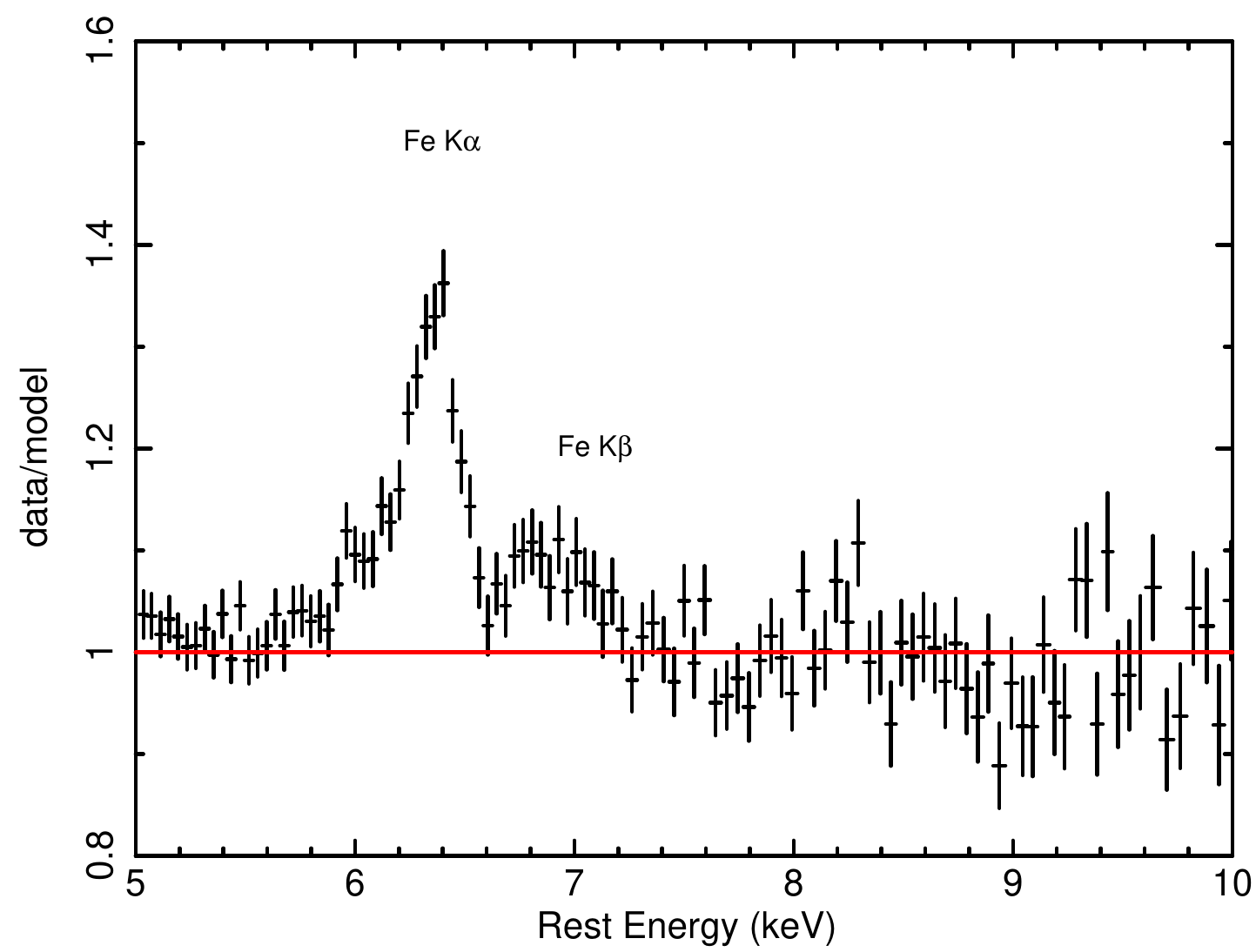}
\caption{{\it Upper}. Broad band ($0.3-79$ keV) residuals when the main continuum is fitted in the $3-5$ and $7.5-10$ spectral regions. The black spectrum represents the EPIC-pn data, FPMA and FPMB data are shown in the same colour, blue for Epoch 1 and green for Epoch 2. {\it Lower}. Zoom on the $5-10$ keV spectral region, with {\it XMM-Newton} (EPIC-pn, black) as a ratio to an absorbed power-law with a photon index $\Gamma=1.86$ and a column density $N_{\rm H}=1.6\times10^{22}$ cm$^{-2}$.}
\label{fig:kalpha_profile}
\end{figure}

\begin{table}
\centering
\caption{Table of the various models used for the hard X-ray spectrum ($E=3-79$ keV).}
\label{tab:3keV}
\begin{tabular}{lcccccccccccccccc}
\hline
Parameter & Model A & Model B\\
\hline
Absorption\\
$N_{\rm H}$ ($10^{22} $cm$^{-2}$) & $1.5\pm0.1$ & $1.7\pm0.2$\\
\hline
Continuum\\
$\Gamma$ & $1.88\pm0.02$ & $1.93\pm0.04$\\
$E_c$ (keV) & $135^{+18}_{-14}$ & $180^{+50}_{-30}$\\
norm$_{\rm PL,obs1}$ ($10^{-2}$ cm$^{-2}$ s$^{-1}$ keV$^{-1}$) & $2.7\pm0.1$ & $2.8\pm0.2$\\
norm$_{\rm PL,obs2}$ ($10^{-2}$ cm$^{-2}$ s$^{-1}$ keV$^{-1}$) & $2.8\pm0.1$ & $3.0\pm0.2$\\
\hline
Slab reflection ({\scriptsize PEXRAV})\\
Abundance & $1^*$ & $-$\\
Fe Abundance & $1^*$ & $-$\\
$\cos\iota$ & $0.45^*$ & $-$ &\\
norm$_{\rm PEX,obs1}$ ($10^{-2}$ cm$^{-2}$ s$^{-1}$ keV$^{-1}$) & $2.2^{+0.3}_{-0.2}$ & $-$\\
norm$_{\rm PEX,obs2}$ ($10^{-2}$ cm$^{-2}$ s$^{-1}$ keV$^{-1}$) & $2.3\pm0.3$ & $-$\\
\hline
Narrow emission lines\\
$E_{\rm K\alpha,n}$ (keV) & $6.36^{+0.02}_{-0.01}$ & $-$\\
EW$_{\rm K\alpha,n}$ (eV) & $49_{-9}^{+7}$ & $-$\\
$\sigma$ & $0^*$ & $-$\\
norm$_{\rm K\alpha,n}$ ($10^{-5}$ cm$^{-2}$ s$^{-1}$ keV$^{-1}$) & $4.6^{+0.7}_{-0.8}$ & $-$\\
$E_{\rm K\beta,n}$ (keV) & $7.06^*$ & $-$\\ 
EW$_{\rm K\beta,n}$ (eV) & $10\pm2$ & $-$\\
\hline
Broad iron line\\
$E_{\rm K\alpha,b}$ (keV)& $6.19^{+0.05}_{-0.06}$ & $-$\\
$\sigma_{\rm K\alpha,b}$ (keV) & $0.27^{+0.06}_{-0.07}$  & $-$\\
EW$_{\rm K\alpha,b}$ (eV) & $56_{-11}^{+9}$ & $-$\\
norm$_{\rm K\alpha,b}$ ($10^{-5}$ cm$^{-2}$ s$^{-1}$ keV$^{-1}$) & $5\pm1$ & $-$\\
\hline
Distant reflection ({\scriptsize XILLVER})\\
Fe Abundance & $-$ & $0.9^{+0.2}_{-0.1}$\\
Inclination  ($^\circ$) & $-$ & $45^{\circ}$ $^*$\\
$\log\xi$ (erg cm s$^{-1}$) & $-$ & $0$ $^*$\\
norm$_{\rm XIL,obs1}$ ($10^{-4}$ cm$^{-2}$ s$^{-1}$ keV$^{-1}$) & $-$ & $2.6^{+0.4}_{-0.2}$\\
norm$_{\rm XIL,obs2}$ ($10^{-4}$ cm$^{-2}$ s$^{-1}$ keV$^{-1}$) & $-$ & $3.0^{+0.5}_{-0.4}$\\
\hline
Relativistic disc reflection ({\scriptsize RELXILL})\\
$R_{in}$ ($R_g$) & $-$ & $56^{+32}_{-19}$\\
$a$ & $-$ & $0^*$\\
$\log\xi$ & $-$ & $2.7\pm0.1$\\
Inclination  ($^\circ$) & $-$ & $45$ $^*$\\
norm$_{\rm REL,obs1}$ ($10^{-5}$ cm$^{-2}$ s$^{-1}$ keV$^{-1}$) & $-$ & $7\pm2$\\
norm$_{\rm REL,obs2}$ ($10^{-5}$ cm$^{-2}$ s$^{-1}$ keV$^{-1}$) & $-$ & $3^{+3}_{-2}$\\
\hline
Velocity shift ({\scriptsize VASHIFT})\\
$v_{\rm shift}$ (km s$^{-1}$) {\it XMM-Newton} & $-$ & $2200\pm300$\\
$v_{\rm shift}$ (km s$^{-1}$) {\it NuSTAR} (Epoch 1) & $-$ & $4300\pm900$\\
$v_{\rm shift}$ (km s$^{-1}$) {\it NuSTAR} (Epoch 2) & $-$ & $5500\pm1000$\\
\hline
$C_{\rm FPMA/pn}$ & $1.39\pm0.01$ & $1.39\pm0.01$\\
$C_{\rm FPMB/pn}$ & $1.42\pm0.01$ & $1.42\pm0.01$\\
\hline
$\chi^2/{\rm dof}$ & $2343/2147$ & $2314/2149$\\
\hline

\end{tabular}
\end{table}


\indent As a first step, we only consider the spectral region between $3$ and $10$ keV for EPIC-pn and both observations of the FPMA and FPMB modules. In order to characterise the shape of the X-ray continuum we also exclude the $5-7.5$ keV region where the iron line is predominant. In all models, we fit the continuum with a simple cut-off power law, absorbed by both Galactic ($N_{\rm H,Gal}=8\times10^{20}$ cm$^{-2}$, \citealp{HI4PI16}) and systemic absorption, using {\scriptsize TBABS} and {\scriptsize ZTBABS}, respectively. We find a photon index $\Gamma=1.86\pm0.01$ and an absorption column density $N_{\rm H}=(1.6\pm0.1)\times10^{22}$ cm$^{-2}$, with an acceptable goodness of fit, $\chi^2/{\rm dof}=604/540$, where dof is the degree of freedom. However, when extrapolated to the energy bands $E<3$ keV and $E>10$ keV, significant residuals arise, due to the presence of a soft X-ray emission and a reflection component, respectively (see upper panel in Fig.~\ref{fig:kalpha_profile}). Moreover, the residuals in the $5-9$ keV energy band show a complex emission including narrow Fe K$\alpha$ and K$\beta$ emission lines, and the presence of a broad component of the Fe K$\alpha$ emission line (see lower panel of Fig.~\ref{fig:kalpha_profile}). At first, we decide to only fit the $3-79$ keV energy band of the spectrum, ignoring the EPIC-pn spectrum below $3$ keV. This choice allows us to find disc parameters with simpler models, not affected by the presence of soft X-ray emission, which then will be used as a starting point for the broad band fits.\\
\indent Our first attempt at fitting the $3-79$ keV energy band is made by considering a cut-off powerlaw at redshift $z=0.00849$ ({\scriptsize ZCUTOFFPL}), and two Fe emission lines. The first emission line represents the typical Fe K$\alpha$ fluorescent line.  We find the centroid energy of the K$\alpha$ line at $E=6.36^{+0.02}_{-0.01}$ keV, which is clear evidence of the presence of the gain issue mentioned in the previous section, as the rest-frame Fe K$\alpha$ line is typically found at $6.4$ keV. The intrinsic width of the narrow K$\alpha$ line was resolved in the {\it Chandra} HETG observation \citep{braito07} to be about 30 eV, which is therefore adopted as a fixed value for our model. We also include a narrow Fe K$\beta$ emission line, with fixed centroid energy $E_{\rm K\beta}=7.06$ keV and intrinsic width 30 eV. The normalisation of the K$\beta$ emission line is assumed to be $13\%$ of that of the K$\alpha$ line \citep[e.g.,][]{palmeri03}. The fit statistics is $\chi^2/{\rm dof}=3185/2152$.\\
\indent  As highlighted by previous observations \citep[e.g.,][]{weaver97,reeves07,braito07,balokovic15,zoghbi17}, this source typically shows a well defined broad iron line in addition to the narrow ones, therefore we add a second Fe K$\alpha$ emission line, this time keeping the width free. We find a line centered at $E=6.19^{+0.05}_{-0.06}$ keV with a width of $\sigma=270^{+60}_{-70}$ eV. The statistic improves to $\chi^2/{\rm dof}=2919/2149$. We note that the blue wing of the broad Fe K$\alpha$ line is degenerate with the possible presence of ionised Fe {\scriptsize XXV} and Fe {\scriptsize XXVI} because of the insufficient spectral resolution.\\
\indent Finally, we also add a reflection component, modelled with {\scriptsize PEXRAV} \citep{magdziarz95}, as pure reflection, i.e. we fix $\mathcal{R}=-1$, with its normalisation allowed to vary independently from the same-epoch power law. We note that all reflection models in this paper are assumed as pure reflection components. Moreover, the continuum parameters of the reflectors ($\Gamma$ and $E_{\rm cut}$) are tied to the ones of the primary continuum. We leave the normalisation of {\scriptsize PEXRAV} in Epoch 1 and Epoch 2 to vary independently. We assume solar abundances, including iron, and assume the default value $\cos \iota=0.45$ for the inclination. The final statistic for this model, which we denote as Model A, is $\chi^2/{\rm dof}=2343/2147=1.09$ (see Tab.~\ref{tab:3keV}).\\
\begin{figure}
    \centering
    \includegraphics[scale=0.35]{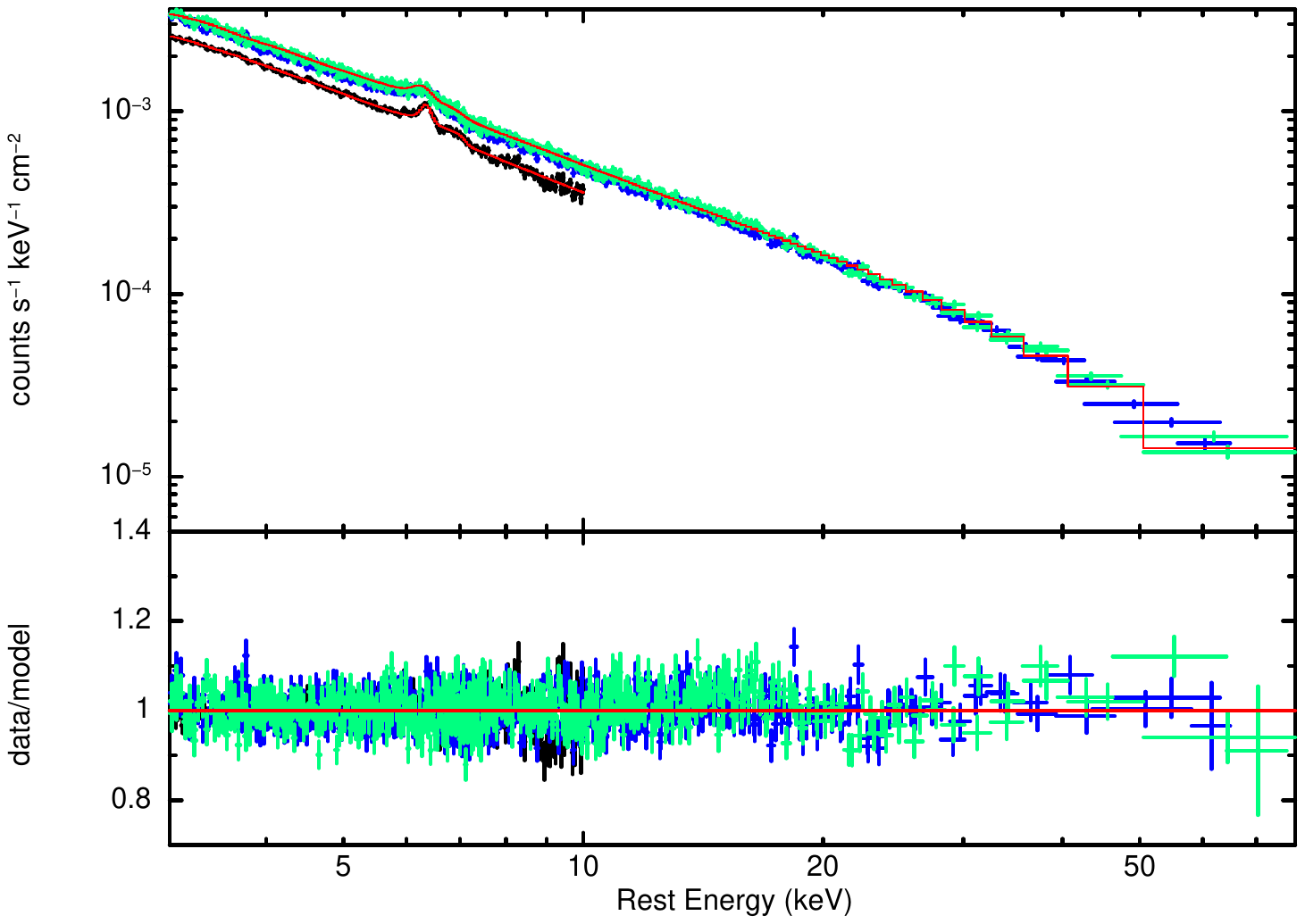}
    \caption{{\it Top}. The $3-79$ keV spectrum analysed here. The black spectrum is the EPIC-pn from {\it XMM-Newton}, while blue and green spectra are the {\it NuSTAR} observations of Epochs 1 and 2, respectively. The two FPM spectra of each observation have been plotted with the same colour and rebinned for visual purposes only. {\it Bottom}. Data-to-model ratio of EPIC-pn and FPM spectra in the $3-79$ keV band, when modelled with Model B.}
    \label{fig:2ra}
\end{figure}
\indent The iron K emission line complex is best fitted by a narrow and a broad component with width $\sigma_{\rm K\alpha,b}=270\pm70$ eV. The latter is most likely produced by reprocessed radiation on the accretion disc, therefore we replace the broad Gaussian line with the accretion disc line emission {\scriptsize RELLINE} \citep{dauser10}. Given that the iron line is not extremely broad like other Seyfert galaxies such as MCG-6-30-15 \citep[e.g.,][]{marinucci14}, we initially adopt a disc radial emissivity index $\beta=-3$ up to $R=1000$ $R_g$, an inclination of $45^{\circ}$ \citep{reeves07} and a non-spinning black hole $a=0$. We also tie the centroid energy to the one of the narrow Fe K$\alpha$ emission line. The result of this fit shows evidence of a truncated disc with inner radius $R_{\rm in}>36\;R_g$, which is consistent with previous results on this source \citep[e.g.,][]{reeves07}. The statistic for this model is $\chi^2/{\rm dof}=2368/2148$. If we assume a maximally spinning black hole ($a=0.998$), the result is unchanged. Indeed, as the disc is truncated, we are not able to infer the ISCO (Innermost Stable Circular Orbit) from the emission line profile and, in turn, the fit does not allow us to measure the black hole spin. Therefore, we will assume $a=0$ in all models in this paper. We also test a different disc radial emissivity profile, assuming a broken power law model with $\beta$ fixed to $-3$ in the outer regions ($R > 30$ $R_g$), while it is left free to vary between $R_{\rm in}= 6$ $R_g$ and $30$ $R_g$. This fit provides $\beta <0$ in the inner regions ($R< 30$ $R_g$), which is still consistent with a truncated disc scenario, indicating a poor contribution of the disc to the line emission for $R< 30$ $R_g$. We will therefore assume an emissivity index of -3 over the whole accretion disc from here on, as it is a typical value for accretion discs above $\sim30\;R_g$ \citep{wilkins12}.\\
\indent However, iron lines likely arise from reflection processes. The presence of both a narrow and a broad iron emission line strongly suggest the presence of two reflection components, a distant one describing the narrow core and a second one representing the reflection on material much closer to the black hole, i.e. the accretion disc. Therefore, we replace the Gaussian and the relativistic lines and the {\scriptsize PEXRAV} reflection model with two reflection continua that also take into account emission lines. For the distant reflector (including the narrow Fe K$\alpha$ line) we use the ionised slab reflector model {\scriptsize XILLVER} \citep{garcia10,garcia13}, that takes into account the most recent atomic data for the iron K fluorescent lines. We fix the value of the inclination to $45^\circ$, and we fit the iron abundance, the ionisation and the normalisation of the slab reflector. We find an upper limit for the ionisation parameter $\log\xi/({\rm erg\; cm\; s^{-1}})<1.3$, which is in agreement with a narrow Fe K fluorescent complex emitted from a distant reflector, for which a poorly ionised medium is expected. The iron abundance is best fitted by $A_{\rm Fe}=0.9^{+0.2}_{-0.1}$.\\
\indent Finally, we interpret the broad iron line as the product of a reflection component from the inner accretion disc, and we model this emission with the relativistic reflection model {\scriptsize RELXILL} \citep{garcia14,dauser14}. This model is able to estimate the inner radius of the accretion disc $R_{\rm in}$, the radial emissivity index, and the spin of the black hole $a$, in addition to every parameter also measured by {\scriptsize XILLVER}. However, since these three parameters are degenerate, we assume an emissivity index of $-3$, as discussed above. We assume a non-spinning black hole with frozen spin $a=0$. The inner radius of the accretion disc is found to be $R_{\rm in}=56^{+32}_{-19}$ $R_g$. We stress that also with this model, choosing different values of the black hole spin does not change any of the other parameters, nor the goodness of fit. This is expected, because, as already anticipated, the region producing the broad iron line does not extend to the ISCO, therefore impeding the measurement of the black hole spin. 
The inclination of the accretion disc is assumed to be aligned with the inclination of the most distant reflector, hence also fixed at $\iota=45^\circ$. We also assume that the iron abundance is the same between the two reflectors. The best-fit ionisation of the accretion disc is $\log\xi/({\rm erg\;cm\;s^{-1}})=2.7\pm0.1$.\\
\indent The final statistic for this model (Model B) is $\chi^2/{\rm dof}=2314/2149=1.07$. The hard band X-ray spectrum and data-to-model ratios relative to Model B are shown in Fig.~\ref{fig:2ra}. A summary of the best-fit values obtained in the hard band is listed in Table~\ref{tab:3keV}.

\subsection{Photoionised plasma in the RGS spectrum}

\label{sec:rgs}

\begin{figure}
    \centering
    \includegraphics[scale=0.35]{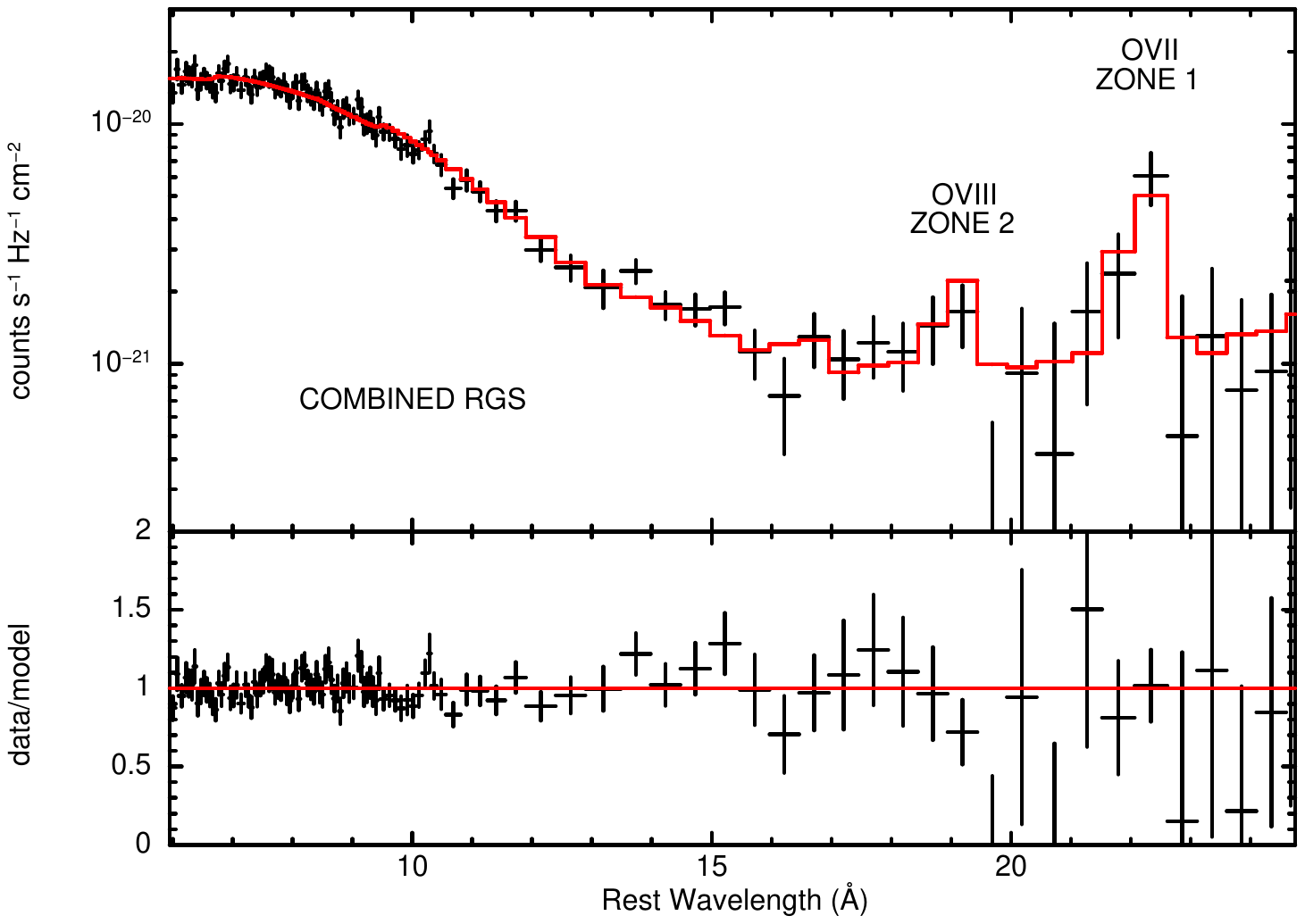}
    \caption{{\it Top}. Combined RGS1+RGS2 spectrum of the analysed data, rebinned here for visual purposes only. The two photoionised plasma components Zone 1 and Zone 2 model the O{\scriptsize VII} and O{\scriptsize VIII} emission components, respectively. {\it Bottom}. Data-to-model ratio of the best-fit model adopted for the RGS spectrum.}
    \label{fig:rgsfit}
\end{figure}

In order to reproduce all the complexities of the broad band data, we analyse the combined RGS spectra of the {\it XMM-Newton} observation. \\
\indent We initially consider the unbinned RGS spectrum. Therefore, we adopt a Cash statistic for our fits \citep{cash79}, as each bins has an insufficient number of counts to adopt the $\chi^2$ statistic \citep[e.g.,][]{kaastra17}. We observe the presence of two prominent emission lines at $\sim19$ \AA $\;$and $\sim22$ \AA. We then select a small spectral region of $\sim100$ channels around these lines and look for Gaussian lines, assuming negligible width ($\sigma=0$). We find a Gaussian line at rest energy $E=(22.08\pm0.01)$ \AA$\;$ ($\Delta C/\Delta {\rm dof}=35/2$) and a second one at $E=(18.98\pm0.01)$ \AA$\;$ ($\Delta C/\Delta {\rm dof}=31/2$). As these emission lines are most likely produced by O {\scriptsize VII} and O {\scriptsize VIII}, respectively, this is a strong hint of the presence of ionized material.\\
\indent We bin the spectrum with wavelength bins of $\Delta\lambda=0.05$ \AA, which samples the resolution of the RGS \citep[e.g.,][]{denherder01}. We still adopt a Cash statistic. We first fit the $\lambda=6-25$ \AA $\,$ spectrum with the superposition of an absorbed and unabsorbed power law. The secondary power law is assumed to be a scattered component from the main one, therefore we tie the two photon indices and we let the normalizations vary independently. We find a photon index of $\Gamma=1.6\pm0.3$, a column density $N_{\rm H}=(2.0\pm0.2)\times10^{22}$ cm$^{-2}$. The goodness of fit is $C/{\rm dof}=441/369$. Significant residuals are present at $\sim 19$ \AA $,$ and $\sim 22$ \AA, as expected from the preliminary analysis described above. We replace the scattering power law with a {\scriptsize CLOUDY} photoionised plasma component \citep{ferland98}, as described in \cite{Bianchi2010a}. We find a ionisation parameter $\log U=1.83^{+0.07}_{-0.04}$ (corresponding to $\log\xi\sim3.3$ erg cm s$^{-1}$ if we adopt an average spectral energy distribution, as described by \citealp{crenshaw12}) and a column density $N_{\rm H}=(4.0^{+0.3}_{-0.5})\times10^{21}$ cm$^{-2}$. The statistic improves to $C/{\rm dof}=420/367$, which improves the fit in the $\sim19$ \AA $\,$ region. However, residuals are still present in the $\sim 22$ \AA $\,$ spectral region, which are not properly fitted by a single photoionised plasma component. Hence, we include a second photoionised plasma component, for which we obtain $\log U=-1.2\pm0.2$ ($\log\xi/{\rm erg\,cm\,s^{-1}}\sim 0.3$) and $N_{\rm H}=2^{+3}_{-1}\times10^{21}$ cm$^{-2}$, improving the fit to $C/{\rm dof}=381/364$. As shown in Fig.~\ref{fig:rgsfit}, the addition of a second photoionised plasma component is able to account for both the $19$ \AA $\,$ and $22$ \AA $\,$ spectral regions. We name the low-ionisation zone as Zone 1 and the high ionisation one as Zone 2. With both the ionised plasma components, the best-fit power law photon index is $\Gamma=1.9^{+0.1}_{-0.2}$ and the absorption column density is $N_{\rm H}=(2.2\pm0.1)\times10^{22}$ cm$^{-2}$.

\subsection{Final broad band model}
\label{sec:broadband}


\begin{figure}
    \centering
    \includegraphics[scale=0.34]{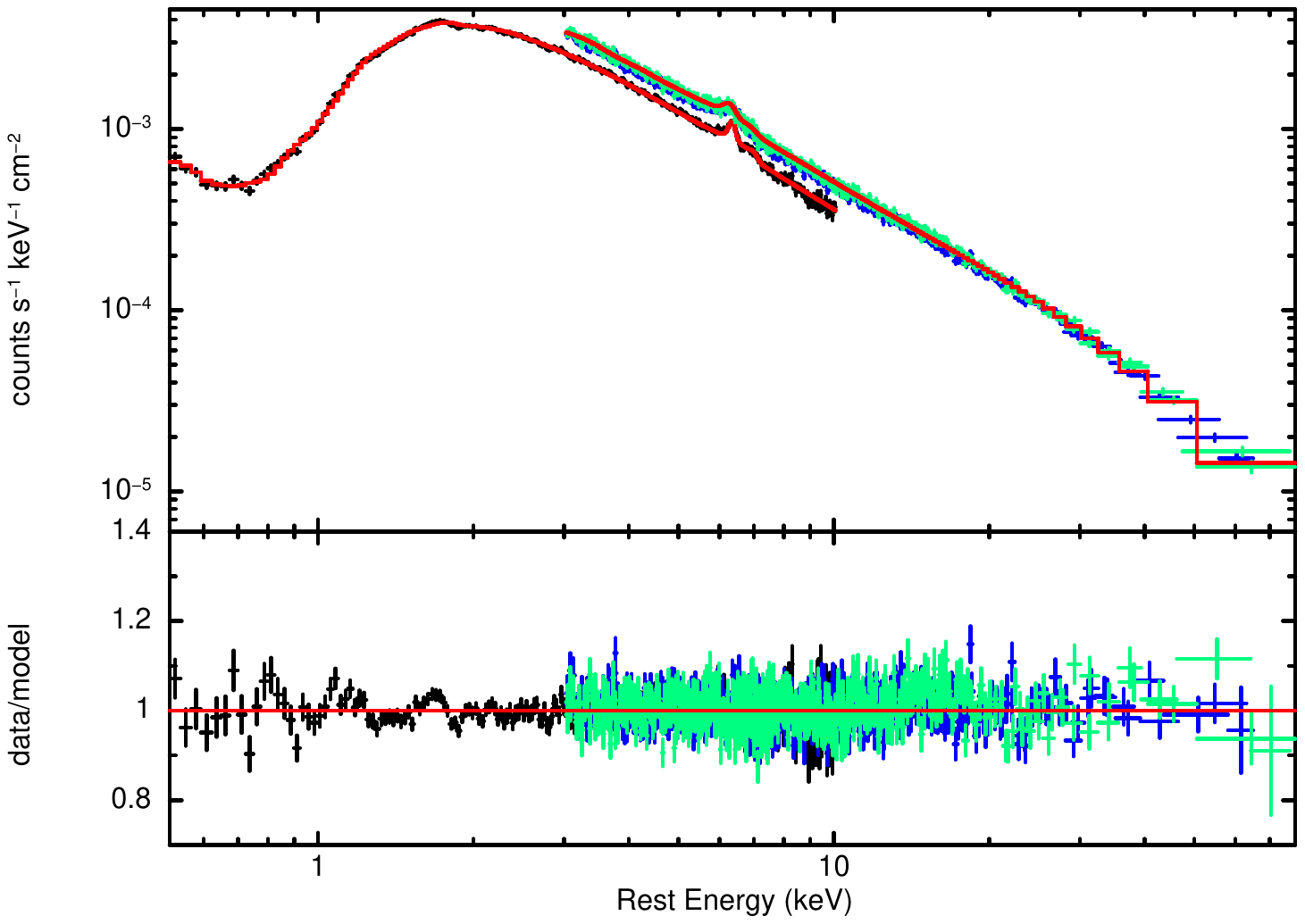}
    \caption{Broad-band {\it XMM-Newton} and {\it NuSTAR} spectra analysed here, fitted with the broad-band best fit model described in Sect.~\ref{sec:broadband}. The black spectrum is the EPIC-pn from {\it XMM-Newton}, while blue and green spectra are the {\it NuSTAR} observations of Epochs 1 and 2, respectively. As in Fig.~\ref{fig:2ra}, the two FPM spectra of each observation have been plotted with the same colour. {\it Bottom}. Data-to-model ratio of EPIC-pn and FPM spectra in the $0.5-79$ keV band.}
    \label{fig:finalspec}
\end{figure}

\begin{table*}
\centering
\caption{Table of the best-fit model values of the broad-band X-ray spectrum ($E=0.5-79$ keV). The goodness of fit for this model is $\chi^2/{\rm dof}=2492/2234$. Given the marginal detection here, the absorption feature at $7.7$ keV is reported in the Table but the goodness of fit refers to the model without the absorption feature. $^{*}$ Frozen parameters. $^{**}$ The internal radius $R_{\rm in}$ is obtained by freezing the inclination $\iota=45^\circ$, while the inclination is obtained by freezing $R_{\rm in}=45R_g$, which would be unconstrained if both the parameters were allowed to vary. Parameters that are tied between Epoch 1 and 2 are only reported in the Epoch 1 column.}
\label{tab:broadtable}
\begin{tabular}{lcccccccccccccccc}
\hline
Parameter & Epoch 1 & Epoch 2\\
\hline
Absorption\\
$N_{\rm H}$ ($10^{22} $cm$^{-2}$) & $1.35\pm0.01$ & - \\
\hline
Central source\\
$\Gamma$ & $1.85\pm0.01$ & -\\
$E_c$ (keV) & $131^{+10}_{-9}$ & - \\
norm$_{\rm PL}$ ($10^{-2}$ cm$^{-2}$ s$^{-1}$ keV$^{-1}$) & $2.44^{+0.03}_{-0.04}$ & $2.57^{+0.04}_{-0.05}$\\
\hline
Distant reflection ({\scriptsize XILLVER})\\
Fe Abundance & $1.4^{+0.3}_{-0.2}$ & -\\
Inclination ($^\circ$) & $45^{\circ}$ $^*$ & -\\
$\log\xi$ (erg cm s$^{-1}$) & $0$ $^*$ & -\\
norm$_{\rm XIL}$ ($10^{-4}$ cm$^{-2}$ s$^{-1}$ keV$^{-1}$) & $1.9^{+0.1}_{-0.2}$ & $2.0^{+0.5}_{-0.4}$\\
\hline
Relativistic disc reflection ({\scriptsize RELXILL})\\
$R_{in}$ ($R_g$) & $40^{+23}_{-16}$ $^{**}$ & -\\
$a$ & $0^*$ & -\\
$\log\xi$ & $2.9^{+0.4}_{-0.2}$ & -\\
Inclination  ($^\circ$) & $41^{+9}_{-11}$ $^{**}$ & -\\
norm$_{\rm REL}$ ($10^{-5}$ cm$^{-2}$ s$^{-1}$ keV$^{-1}$) & $4\pm1$ & $2\pm1$\\
\hline
Photoionised emission Zone 1 ({\scriptsize CLOUDY}, EPIC-pn)\\
$\log U$ & $>-2$ & -\\
$N_{\rm H}$ (cm$^{-2}$) & $(5\pm4)\times10^{21}$ & -\\
\hline
Photoionised emission Zone 2 ({\scriptsize CLOUDY}, EPIC-pn)\\
$\log U$ & $2.2^{+0.1}_{-0.3}$ & -\\
$N_{\rm H}$ (cm$^{-2}$) & $(6\pm1)\times10^{22}$ & -\\
\hline
\hline
Fluxes\\
$F_{0.5-2{\rm \;keV}}$ ($10^{-12}$ erg s$^{-1}$ cm$^{-2}$) & $8.19\pm0.04$ & -\\
\\
$F_{2-10{\rm \;keV}}$ ($10^{-11}$ erg s$^{-1}$ cm$^{-2}$) & $7.40^{+0.02}_{-0.01}$ & $7.92\pm0.02$\\
\hline
Unabsorbed luminosity\\
$L_{2-10{\rm \;keV}}$ ($10^{43}$ erg s$^{-1}$) & $1.24\pm0.01$ & $1.30\pm0.02$\\
\hline

\end{tabular}
\end{table*}

We now consider the broad band ($E=0.5-79$ keV) EPIC-pn and {\it NuSTAR} data.\\
\indent We use the RGS results described in Sect.~\ref{sec:rgs} to model the soft X-ray emission, therefore we add to the hard energy band model described in Sect.~\ref{sec:hardband} two Galaxy-absorbed photoionised plasma components with ionisation and column density values fixed to the ones obtained with the RGS fit, allowing the two normalisations to vary. The {\scriptsize XSPEC} expression of the broad band model is therefore\\

\noindent {\tt TBabs*zTBabs*(zcutoffpl+vashift*(relxill+xillver))+}\\
{\tt +TBabs*(cloudy1+cloudy2).}\\

The statistic of this model is $\chi^2/{\rm dof}=2579/2238$. There are some residuals in the $E<2$ keV energy band. Given the best signal-to-noise ratio of EPIC-pn data with respect to the RGS, we free the two photoionized components, obtaining a better fit ($\chi^2/{\rm dof}=2492/2234\simeq1.11$, corresponding to $\Delta\chi^2/\Delta{\rm dof}=87/4$). The low ionization Zone 1 has an unconstrained ionisation ($\log U>-2$) with a column density $N_{\rm H}=(5\pm4)\times10^{21}$ cm$^{-2}$, consistent with the RGS result. The high ionization Zone 2 has a ionisation parameter of $\log U=2.2^{+0.1}_{-0.3}$, consistent with the RGS, while the column density is $N_{\rm H}=(6\pm1)\times10^{22}$ cm$^{-2}$. A few residuals are still present, as shown in Fig.~\ref{fig:finalspec}, possibly arising from emission lines at $E\simeq0.80$ keV $E\simeq1.16$ keV and $E\simeq1.68$ keV. These lines are not present in the RGS spectrum, therefore their nature is not clear, but they are unlikely to arise from the source. Moreover, the most significant of these lines has a centroid energy of $E=1.67^{+0.01}_{-0.03}$ keV, which is not a known emission line in AGN. However, investigating the nature of these lines is beyond the scope of this paper.\\
\indent We find a column density of the absorber of $N_{\rm H}=(1.35\pm0.01)\times10^{22}$ cm$^{-2}$. The main power law parameters are a photon index of $\Gamma=1.85\pm0.01$, with a cut-off energy $E_{\rm cut}=131^{+10}_{-9}$ keV. The best-fit values of the broad band model are summarised in Table~\ref{tab:broadtable}.

\section{Discussion}
\subsection{Continuum}
\label{sec:discocont}

\begin{figure}
    \centering
    \includegraphics[scale=0.35]{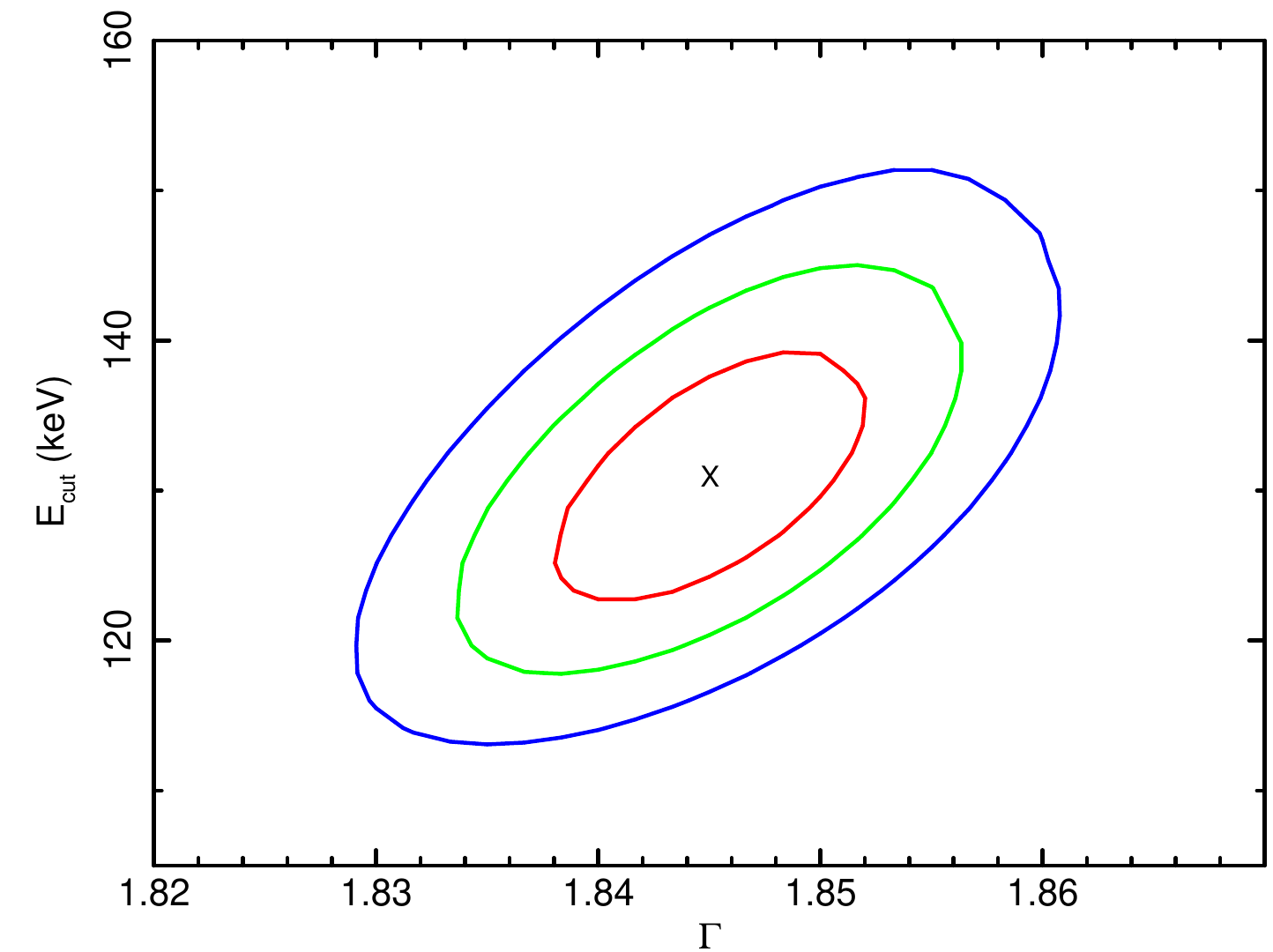}
    \caption{Contour plot $E_{\rm cut}-\Gamma$ for the broad-band model. The red, green and blue lines represents the $68\%$ ($1\sigma$), $95\%$ ($2\sigma$) and $99.7\%$ ($3\sigma$) confidence levels, while the best fit value is marked by an X.}
    \label{fig:ecut_gamma_broad}
\end{figure}

As shown in Sect.~\ref{sec:broadband}, the X-ray continuum is well fitted by a photon index of $\Gamma=1.85\pm0.01$ and a cut-off energy $E_{\rm cut}=131^{+10}_{-9}$ keV. Both the photon index and the cut-off energy are consistent with previous {\it NuSTAR} results of this AGN \citep{balokovic15}, where a cut-off energy of $E_{\rm cut}=116^{+6}_{-5}$ keV was found.  In Fig.~\ref{fig:ecut_gamma_broad} the $\Gamma-E_{\rm cut}$ contour plot for $1\sigma$, $2\sigma$, and $3\sigma$ confidence levels is shown. We also investigate the possible variability of the photon index and cut-off energy, which was suggested in past observations of MCG-5-23-16 \citep{zoghbi17}. If we let $\Gamma$ and $E_{\rm cut}$ to vary independently between Epoch 1 and Epoch 2, keeping the two parameters tied between {\it XMM-Newton} and {\it NuSTAR} in Epoch 1, we obtain identical photon indices. In Epoch 1 the cut-off energy is $E_{\rm cut,1}=113^{+12}_{-10}$ keV, while for Epoch 2 the cut-off energy is $E_{\rm cut,2}=140^{+18}_{-16}$ keV, which are consistent at the $90\%$ confidence value.\\
\indent The photon index and the cut-off energy are tightly related to the coronal electron temperature and optical depth \citep[e.g.,][]{petrucci01,middei19}, which could be investigated with Comptonisation models. We estimate the temperature $kT$ and the optical depth $\tau$ of the corona of MCG-5-23-16 by replacing the cut-off power law continuum with the physical model {\scriptsize COMPPS} \citep{poutanen96} assuming a slab geometry (i.e. we set the parameter {\tt geom} to $1$). We assume a black body temperature of the seed photons of $10$ eV and we find $kT=26\pm1$ keV and $\tau=1.25\pm0.05$, consistent at the $90\%$ confidence level with \cite{balokovic15} and \cite{marinucci22}. If we explore a spherical geometry (i.e. {\tt geom}=0) we obtain $kT=28^{+6}_{-5}$ keV and $\tau=4.7^{+0.5}_{-1.9}$.

\subsection{Reflection}

The reflection spectrum of MCG-5-23-16 is comprised of two different components. The distant reflecting medium, also responsible for the narrow core of the Fe K$\alpha$ emission line, is likely associated with the obscuring torus. The inner reflecting region, producing the broad component of the emission line, is instead likely associated with the accretion disc.\\
The iron abundance is kept tied between the two reflectors and it is $A_{\rm Fe}=1.4^{+0.3}_{-0.2}$. We initially keep the inclination fixed at $45^\circ$, and we assume a non-spinning black hole, i.e. $a=0$. We note that if we assume a different value of the black hole spin, such as a maximally spinning black hole with $a=0.998$, our best-fit values are left almost unchanged at the $90\%$ confidence level. We also assume an radial emissivity index of $-3$. The disc turns out to be truncated, with an inner radius of $R_{\rm in}=40^{+23}_{-16}$ $R_g$, in agreement with the past observations of this AGN. The ionisation parameter of the disc reflector is $\log\xi/{\rm (erg\,cm\,s^{-1})}=2.9^{+0.4}_{-0.2}$, while that of the distant reflector is consistent with a neutral or weakly ionised medium. If we fix the inner radius to $45$ $R_g$ and let the inclination free we obtain $\iota=41^\circ\;^{+9}_{-11}$. However, the two parameters are strongly degenerate, and if we allow them both to vary we do not constrain the inner radius, obtaining $R_{\rm in}>6$ $R_g$ with an inclination of $\iota=31\pm10^\circ$. Although this inclination value is consistent with the one obtained assuming a truncated disc, extremely low inclinations are highly unlikely given the persistent moderate obscuration, which is always observed in MCG-5-23-16. Indeed, a persistent obscuration with values of the column density of a few $10^{22}$ cm$^{-2}$, as in the case of MCG-5-23-16, is likely due to the line of sight crossing near the edge of the torus. Assuming that the torus and the accretion disc are aligned, it is unlikely that small face-on inclinations are able to describe the observed lack of absorption variability. Alternatively, the column density may be due to a galactic dust lane, but a more face-on inclination may still appear less likely given its type 1.9 optical classification \citep[e.g.,][]{veron80}.\\
\indent We also test if a maximally spinning black hole, i.e. $a=0.998$, changes any result when both the inner radius and inclination are let free. We find the same result, i.e. an unconstrained inner radius and a smaller inclination, with the lower value of the inner radius set to $R_{\rm in}>1.7$ $R_g$.

\subsection{A possible absorption complex in the Fe K band}

Some absorption residuals may be also present in the EPIC-pn data at the Fe K band. We test the possible presence of absorption lines at $\sim7-10$ keV with Gaussians. The inclusion of a line at $\sim7.7$ keV improves the fit by $\Delta\chi^2/\Delta{\rm dof} = 15/3$. If we simulate 1000 EPIC-pn spectra adopting the best-fit model, without any absorption line, we find 55 spectra for which a spurious absorption line in the $7-9$ keV is found with $\Delta\chi^2/\Delta{\rm dof}\geq15/3$ \citep[see e.g.,][for details on this procedure]{markowitz06}, which implies a $\lesssim95\%$ confidence level. While this is a marginal detection, an absorption complex was already found by \cite{braito07}, who interpreted a feature at $\sim7.7$ keV as an absorption line from Fe {\scriptsize XXVI}, outflowing with velocity $v\sim0.09c$. Therefore, purely in order to compare with the previous detection of the line, we report here its best-fit parameters.\\
\indent We fit the absorption trough at $\sim7.7$ keV with a simple Gaussian line, with width $\sigma$ left free to vary. We find an unconstrained width and therefore we fix $\sigma=0$. We let the normalisations in Epoch 1 and 2 free to vary independently. We find a centroid energy of $E=7.74^{+0.05}_{-0.06}$ keV, with an equivalent width of EW$_{1}=19\pm5$ eV in Epoch 1 and EW$_{2}<5$ eV in Epoch 2. Therefore, if a line is indeed present, it would be two times weaker than previously observed in \cite{braito07}, where an equivalent width of EW$\sim50$ eV was found. This would be most likely due to a change in column density of the wind. However, the centroid energy is consistent with the one recovered in the past, which would imply an outflowing velocity of $v/c=0.10\pm0.01$, if interpreted as outflowing Fe {\scriptsize XXVI}. However, we stress that the line is not indeed detected, and only mentioned here to compare it with previous 2005 observations of this AGN.

\section{Summary and conclusions}
\label{sec:theend}

We have presented the analysis of one {\it XMM-Newton} and two {\it NuSTAR} observations of MCG-5-23-16, one of which contemporaneous with {\it XMM}. This is the first contemporaneous broad-band observation of this AGN with the two telescopes. Both {\it NuSTAR} epochs are also contemporaneous with {\it IXPE} observations, for which upper limits on the polarisation degree were observed \citep{marinucci22,tagliacozzo23}. We summarise our results in the following
\begin{itemize}
\item We find a well constrained value of the cut-off energy at $E_{\rm cut}=131^{+10}_{-9}$ keV. This value is consistent with previous results of {\it NuSTAR} analyses on this AGN and it represents an average value of the cut-off energies typically found in AGN \citep[e.g.,][]{tortosa18,kamraj22}. No variability of the cut-off energy is detected between Epoch 1 and Epoch 2. Assuming a slab (spherical) geometry of the corona, we find a coronal temperature of $kT=26\pm1$ keV ($kT=28^{+6}_{-4}$) and an optical depth $\tau=1.25\pm0.05$ ($\tau=4.7^{+0.5}_{-0.9}$), largely in agreement with previous results on this AGN.
\item The spectrum shows evidence of the presence of a broad line, produced by the the reflection of the primary component on a truncated accretion disc, with an inner radius of $R_{\rm in}=40^{+23}_{-16}$ $R_g$. 
\item The inclination is $\iota=41^\circ$$^{+9}_{-10}$, which is consistent with a persistent moderate obscuration as that observed for MCG-5-23-16. Indeed, we find $N_{\rm H}\sim1.3\times10^{22}$ cm$^{-2}$, roughly constant in time scales of decades. A persistent obscuration with such column density, if due to the absorption by the torus, would imply that the absorption occurs at the edge of the torus itself, and an inclination of $\sim45^\circ$, consistent with our result, would be required. \cite{tagliacozzo23} showed that, even though the polarisation degree is not constrained at the $99\%$ confidence level, if we assume the inclination obtained in the present paper a wedge corona \citep[e.g.,][]{poutanen18} is favoured, as in the case of NGC 4151 \citep{gianolli23}.
\item The RGS spectrum highlights the presence of two photoionised plasma components. When the EPIC-pn is fitted, the first component, denoted as Zone 1, is characterised by a ionisation parameter $\log U>-2$ and a column density of $N_{\rm H}\sim 5\times10^{21}$ cm$^{-2}$, which is responsible of the O {\scriptsize VII} emission in the RGS spectrum. A second component, Zone 2, is characterised by a larger ionisation parameter ($\log U\sim 2.2$) and column density of $N_{\rm H}\sim 6\times10^{22}$ cm$^{-2}$, which is instead responsible for the more ionised O {\scriptsize VIII} emission line.
\item We also report on the marginal detection of a possible absorption line with centroid energy $E~\sim7.74$ keV. If confirmed, the line could be interpreted as blue-shifted outflowing Fe {\scriptsize XXVI}, which would imply a wind velocity of $v/c\sim 0.1c$. However, we must stress that the line is only marginally detected, due to a smaller column density with respect to the previous detection of the line \citep{braito07}, and we only report the possible absorption feature to compare it with its prior detection with {\it XMM-Newton}.
\item Overall, the X-ray spectrum of MCG-5-23-16 appears to be remarkably stable over the years. In fact, only moderate variations in flux and column density are observed, while the properties of the disc, such as the inner radius inferred from the broad iron line, do not appear to significantly change compared to past observations with {\it Suzaku} \citep{reeves07}, {\it XMM-Newton} \citep{braito07}, and {\it NuSTAR} \citep{balokovic15,zoghbi17}. Furthermore, the X-ray coronal properties also appear consistent, as compared between the observations presented here and the past {\it NuSTAR} pointings.
\end{itemize}
Future high-resolution observations of MCG-5-23-16 with the microcalorimeter Resolve on board the upcoming X-ray mission {\it XRISM} \citep{xrism20} will allow us to observe the X-ray sources with unprecedented spectral resolution. XRISM observations will be able to resolve the various components of the iron line profile, such as the possible presence of ionized iron lines. Indeed, this will be groundbreaking in determining the disc parameters from the broad iron line with much larger accuracy, in particular the inner radius $R_{\rm in}$ and the inclination of the accretion disc, as the width of the narrow core and possibly any contribution from the broad line region will be measured with unprecedented accuracy, clearly separating it from the disc-reflection component.\\
Additionally, the future enhanced X-ray, Timing and Polarimetry mission \citep[{\it eXTP},][]{zhang19,derosa19}, will be able to perform simultaneous spectral-timing polarimetry measurements, allowing us to put strong constraints on the emission and reflection properties of accreting SMBHs.

\section*{Acknowledgements}

The authors thank the referee for significantly improving the quality of this paper with useful comments and suggestions. RS and ADR acknowledge financial support from the agreements ASI-INAF eXTP Fase B - 2020-3-HH.1-2021 and   ASI-INAF n.2017-14-H.O. JR and VB acknowledge financial support through NASA grants 80NSSC22K0474, 80NSSC22K0003 and 80NSSC22K0220. POP aknowledges financial support from the High Energy programme (PNHE) of the Scientific Research National Center (CNRS) and the French Space Agency (CNES). SB acknowledges support from PRIN MUR 2017 ``Black hole winds and the baryon life cycle of galaxies: the stone-guest at the galaxy evolution supper'', and from the European Union Horizon 2020 Research and Innovation Framework Programme under grant agreement AHEAD2020 n. 871158. This research has made use of data and software provided by the High Energy Astrophysics Science Archive Research Center (HEASARC), which is a service of the Astrophysics Science Division at NASA/GSFC and the High Energy Astrophysics Division of the Smithsonian Astrophysical Observatory. This research has made use of the NuSTAR Data Analysis Software (NUSTARDAS), jointly developed by the ASI Space Science Data Center (SSDC, Italy) and the California Institute of Technology (Caltech, USA). The research is partly based on observations obtained with XMM-Newton, an ESA science mission with instruments and contributions directly funded by ESA Member States and NASA

\section*{Data availability}

The {\it XMM-Newton} data underlying this article are subject to an embargo and will be publicly available from 2023-07-08 from the XMM-Newton science archive (\url{http://nxsa.esac.esa.int/}). The {\it NuSTAR} data are publicly available on the online archive (\url{https://heasarc.gsfc.nasa.gov/docs/nustar/nustar\_archive.html}).



\bibliographystyle{mnras}
\bibliography{biblio} 







\bsp	
\label{lastpage}
\end{document}